\documentclass[reprint,amsmath,amssymb,aps,prb]{revtex4-2}
\usepackage{stix2}
\usepackage{graphicx}
\usepackage{bm}
\usepackage{hyperref}       
\usepackage{epstopdf}
\usepackage{float}
\usepackage{subfigure}
\usepackage{threeparttable}
\usepackage{booktabs}
\usepackage{xcolor}
\usepackage{multirow}

\begin{document}

\title{Symmetry-adapted graph neural networks for constructing molecular dynamics force fields}

\author{Zun \surname{Wang}$^1$}
\author{Chong \surname{Wang}$^2$}
\email{ch-wang@outlook.com}
\author{Sibo \surname{Zhao}$^1$}
\author{Shiqiao \surname{Du}$^1$}
\author{Yong \surname{Xu}$^{1,4,5}$}
\author{Bing-Lin \surname{Gu}$^{1,3}$}
\author{Wenhui \surname{Duan}$^{1,3,4}$}
\email{duanw@tsinghua.edu.cn}
\affiliation{$^1$State Key Laboratory of Low Dimensional Quantum Physics and Department of Physics, Tsinghua University, Beijing, 100084, China\\
$^2$Department of Physics, Carnegie Mellon University, Pittsburgh, Pennsylvania 15213, USA\\
$^3$Institute for Advanced Study, Tsinghua University, Beijing 100084, China \\
$^4$Frontier Science Center for Quantum Information, Beijing 100084, China\\
$^5$RIKEN Center for Emergent Matter Science (CEMS), Wako, Saitama 351-0198, Japan}

\date{\today}

\begin{abstract}
Molecular dynamics is a powerful simulation tool to explore material properties. Most of the realistic material systems are too large to be simulated with first-principles molecular dynamics. Classical molecular dynamics has lower computational cost but requires accurate force fields to achieve chemical accuracy. In this work, we develop a symmetry-adapted graph neural networks framework, named molecular dynamics graph neural networks (MDGNN), to construct force fields automatically for molecular dynamics simulations for both molecules and crystals. This architecture consistently preserves the translation, rotation and permutation invariance in the simulations. We propose a new feature engineering method including higher order contributions and show that MDGNN accurately reproduces the results of both classical and first-principles molecular dynamics. We also demonstrate that force fields constructed by the model has good transferability. Therefore, MDGNN provides an efficient and promising option for molecular dynamics simulations of large scale systems with high accuracy. 
\end{abstract}

\maketitle


\section{\label{sec:level1}Introduction}

Molecular dynamics (MD) is an indispensable simulation method in physics, chemistry, biology and material science. While current first-principles MD can feasibly compute material systems of small sizes, the computational cost is formidable for large scale simulations~\cite{car1985unified}. On the other hand, the computational cost for classical MD is typically smaller than first-principles MD by orders of magnitude, but the accuracy depends heavily on the precision of force fields, which describe the interactions between atoms. Traditional construction of force fields starts with predetermined forms of force fields and then the parameters are fitted to free energies of different atomic configurations~\cite{vanommeslaeghe2010charmm,jorgensen1996development,wang2004development}. This process relies on expert knowledge of computational quantum chemistry and generally cannot be automated.

Recently, machine learning methods emerge as an efficient tool for handling tasks where human labor was thought to be indispensable. Especially, with the tremendous expressive power~\cite{hornik1989multilayer,hornik1991approximation} and generalization ability of deep neural networks, reconstructing accurate force fields with fewer data points becomes possible. Several successful schemes that reproduce force fields by machine learning have been proposed, including the Behler-Parrinello neural networks~\cite{behler2007generalized}, the gradient-domain machine learning~\cite{chmiela2017machine}, the deep tensor neural networks~\cite{schutt2017quantum} and deep potential molecular dynamics~\cite{han2017deep,zhang2018deep,wang2018deepmd}. The free energy of a system is manifestly invariant under rotation, translation and permutation within the atoms of the same kind. This point, although trivial to human beings, is not transparent to the neural networks if the Cartesian coordinates are used directly as inputs. To optimize the outcome of force fields constructed by machine learning, it is necessary to expose all the known symmetries to the neural networks. For this purpose, the Behler-Parrinello neural networks~\cite{behler2007generalized} construct a set of many-body symmetry functions of coordinates as the input, which can only handle a few number of distinct elements. The loss function of the Behler-Parrinello neural networks~\cite{behler2007generalized} merely includes the contribution of energies, which usually makes the training process of the model trapped into a local minimum~\cite{ciccotti2008projection,schutt2017schnet}. The gradient-domain machine learning~\cite{chmiela2017machine} and its variant (the symmetric gradient domain machine learning~\cite{chmiela2018towards}) are kernel-based statistical learning models, in which the coordinates are mapped to the ordered eigenvalues of the Coulomb matrix. The input features of the deep tensor neural networks~\cite{schutt2017quantum}, which is proposed for small molecules, are charges and distances expanded in a Gaussian basis. Deep potential molecular dynamics (DeePMD)~\cite{han2017deep,zhang2018deep,wang2018deepmd,zhang2018endtoend} has been developed as a systematic pipeline to reduce data preprocessing, but the coordinates still need to be transformed into local spherical coordinate frames and ordered with some rules to preserve all the symmetries. The reason of heavy data preprocessing in the previous models is that the graph-structured data could not be processed directly by fully connected neural networks or statistical learning methods. A straightforward and intuitive method of characterizing the atom configurations is still lacking.


Recently, graph neural networks (GNNs) have shown great potential in learning from graph-structured data, and has become an increasingly important tool in machine learning~\cite{scarselli2008graph,hamilton2017representation,bruna2013spectral,zhou2018graph,wu2020comprehensive,zhang2020deep}. SchNet~\cite{schutt2017schnet} is a GNN-based framework whose inputs are the embeddings of charges and the Gaussian expansion of distances. DimeNet~\cite{klicpera2019directional} is another novel GNN model whose features are inspired by solutions of time-independent Schr\"{o}dinger equation. Innovatively, DimeNet~\cite{klicpera2019directional} explicitly introduces angular information to GNN with directed graphs. However, these two models are only applied to small molecules. The tensor embedded atom network (TeaNet)~\cite{takamoto2019teanet} is a GNN-based framework proposed to construct force fields for crystals. Similar to iterative relaxation, TeaNet repeats its interaction block several times to predict the ground state energies. Since the interaction blocks share the same parameters, TeaNet is actually an original GNN~\cite{scarselli2008graph} framework in analogy with recursive neural networks~\cite{frasconi1998general,sperduti1997supervised,hagenbuchner2003self}, which probably tends to suffer from over-smooth issues~\cite{li2018deeper}.

In this work, we propose a symmetry-adapted GNN framework, named Molecular Dynamics Graph Neural Network (MDGNN), to construct accurate force fields for both molecules and crystals. Compared to the traditional GNN which preserves the permutation invariance, MDGNN additionally imposes restrictions to respect translation and rotation invariance. With a new feature engineering method proposed to introduce higher order effects, we show this architecture can reproduce the potential-energy surface (PES) up to the accuracy of first-principles methods for both molecules and crystals. The transferability of MDGNN-constructed force fields is demonstrated by simulating materials that MDGNN was not trained with. Compared to first-principles MD and even classical MD, MDGNN is computationally more efficient. With minimal data preprocessing, MDGNN automates the construction of force fields and is a computationally efficient tool to explore material properties.

\begin{figure*}
    \centering
    \includegraphics[width=1.0\textwidth]{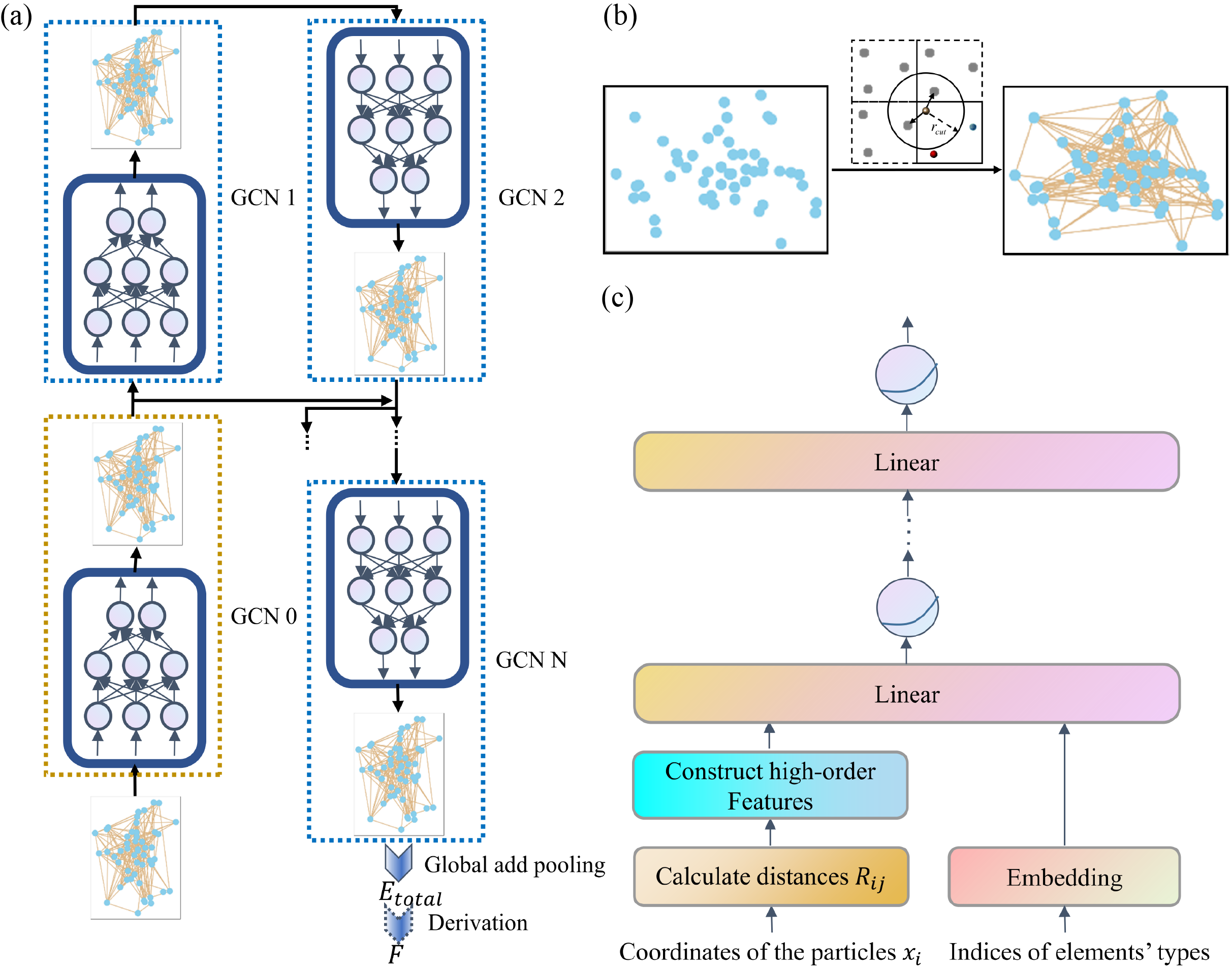}
    \caption{The architecture diagram of MDGNN. (a) The overall structure of the model. The input of the model is a molecule graph or a crystal graph transformed from a graph-structured frame of molecular dynamics trajectory. Processed by a translation- and rotation-preserved graph convolutional network (GCN) layer named GCN0 and $N$ GCN layers, the node features of the output graph represent the local energies of the center atoms. These features will be element-wise added after every two GCN layers, same as residual networks. The total energy is obtained by summing over all the atomic features directly. The specific local energies are not required and the model is trained end-to-end to predict the total energy of the supercell. To obey the energy-conservation rule, the forces on each atom are the derivatives of the total energy with respect to the corresponding coordinates. (b) The crystal graph is built with periodic boundary condition and minimum image convention. For an isolated molecule, 15 \AA~ vacuum layers are built for each dimension, and then the molecule graph is equivalent to a crystal graph. (c) The details of GCN0 layer which preserves translation and rotation invariance. The inputs of GCN0 are the embeddings of atomic types and high-order features of $R_{ij}$ of atomic pairs. After several multi-layer perceptrons (MLPs), higher dimensional node features are obtained.}
    \label{fig:1}
\end{figure*}

\section{Results}

\subsection{Architecture\label{sec:architecture}}
With atoms identified as vertices and interatomic interactions identified as edges, GNN can naturally describe a collection of atoms and all the symmetries mentioned previously can be automatically enforced. In this work we focus on undirected graphs $\mathcal G$. The goal of the GNN is to achieve the hidden state embeddings of each node by graph perception. With the neighborhoods of the node $i$ passing their messages to the center, the hidden states of the node $i$ will contain the information from the neighborhoods. Gilmer {\it et al.}~\cite{gilmer2017neural} proposed the recapitulative formula of the GNNs, 
\begin{equation}
    \mathbf{x}_i^{(k)} = \gamma_{\mathbf{\Theta}}^{(k)} \left(\mathbf{x}_i^{(k-1)},\square_{j\in\mathcal{N}_{(i)}}\phi_{\mathbf{\Theta}}^{(k)}\left(\mathbf{x}_i^{(k-1)},\mathbf{x}_j^{(k-1)},\mathbf{e}_{ij}^{(k-1)}\right)\right)
\end{equation}
where the superscript $k$ represents the $k$-th GNN layer, and $\mathbf{x}_i$ and $\mathbf{x}_j$ are the features of node $i$ and node $j$, respectively. $\mathbf{e}_{ij}$ denotes the feature of the edge connecting nodes $i$ and $j$. $\phi_{\mathbf{\Theta}}$ and $\gamma_{\mathbf{\Theta}}$ are message passing function and updating function, respectively. The specific forms of these functions could be expressed as multi-layer perceptrons (MLPs). $\mathcal{N}(i)$ denotes the neighborhoods of node $i$ and $\square$ is the aggregation operator (e.g. sum or average), which ensures the permutation invariance of the graph data. In fact, the formula implements a spatially convolutional operation on the graph, thus the architecture is termed Graph Convolutional Network (GCN)~\cite{kipf2016semi}. This formula should be viewed as a two-step continuous procedure, i.e., message passing step and updating step. In the message passing step, neighboring nodes pass their information to the central node (the convolution center). The central node then performs an aggregation operation on the informing information and correspondingly update its hidden state. The final hidden states of each node are obtained after several GNN layers and are used to carry out node-level tasks. In our model, a graph-level task is conducted with a final readout layer~\cite{ying2018hierarchical,zhang2018end}.

The architecture of the model is drawn schematically in Fig. \ref{fig:1} (a). The input of the framework is a graph transformed from a frame of the molecular dynamics trajectory. Periodic boundary condition will always be adopted, even for isolated molecules, where vacuum layers are inserted between repeated images. The inputs of the computational graph are only the embeddings of atomic type and the coordinates without any data preprocessing. The neural network is expected to predict the potential energy of the corresponding configuration, which should satisfy translation, rotation and permutation invariance. As mentioned above, GNNs could self-consistently enforce permutation invariance. To further ensure translation and rotation invariance, the distances $R_{ij}$ between the atoms are computed as the edge features. It must be emphasized that due to the periodic boundary condition, the distance $R_{ij}$ is calculated by taking the minimum of the distances between node $i$ and the periodic images of node $j$. To save the computational cost, only interatomic interactions within a fixed truncation radius $r_c$ will be calculated. To ensure the distance between two atoms is unique, the minimum image convention requires that the shortest lattice parameter of the supercell is larger than $2 r_c$. The input edge features of GCN1 layer which describe atomic positions is the concatenation of $\sin[l (R_{ij}/r_c)^m/m!]$ and $\cos[l (R_{ij}/r_c)^m/m!]$, where $l = 1, 2, \dots, L$ and $m=1, 2, \dots, M$. These features include high order terms of $R_{ij}$ and take value from $-1$ to 1. Besides the above features, the embeddings~\cite{gal2016theoretically} of the two relevant atomic types are also concatenated as the inputs of the next block of the neural network. There are other methods that could be used to distinguish the atomic types, such as charges or masses. We find that including these features does not improve MDGNN's performance. This means that these features only distinguish the types of atoms while the specific parameters of force fields will be learned by training and the principal contributions to the interactions between atoms are the distances. We choose summation as the aggregation operation to ensure the permutation invariance. Note that summation is the only operation that makes sense since forces and energies are additive. The above process can be viewed as a GCN layer with physically motivated constraints, which is denoted as GCN0 layer in Fig. \ref{fig:1} (a). 
In a single GCN layer, messages from a node's $1$-hop neighborhood will be aggregated to center~\cite{kipf2016semi}. Similarly, after $k$ GCN layers, the central node will compute messages within $k$-hop distance~\cite{xu2018representation}. The network comprises several different graph convolutional network's convolutional~\cite{kipf2016semi} layers, and the last GCN layer outputs local energies. Nonlinear activations are applied to all but the final convolutional layer. Note that the hidden state embeddings are element-wise added every two GCN layers, similar to residual connections~\cite{he2016deep}, which allows GCNs go deeper without suffering from over-smoothing~\cite{li2018deeper}. The global add pooling simply adds the local potential energies of all atoms, yielding the potential energy of the supercell. To ensure energy conservation, we obtain forces by taking the gradients of the local energies with respect to the positions of the input coordinates through automatic differentiation of the network and preprocessing operations. This practice is only justified when only conservative forces are involved in the dynamics. Notable exceptions include magnetic field, for which MDGNN cannot be used to perform MD simulations.

\subsection{Training details}

The dataset comprises frames from MD simulations: atom types, atomic coordinates, potential energies and forces at multiple time steps. Atomic types and interatomic distances are the only inputs to the model. The dataset which is transformed to molecular or crystal graphs, is randomly shuffled and split into training set, validation set and test set with a certain ratio. The mean value of the energies of the dataset is subtracted from the energies of all frames. With only atomic embeddings and interatomic distances as initial features, the symmetry-adapted MDGNN preserves rotation and translation symmetry. MDGNN is programmed with the PyTorch~\cite{NEURIPS2019_9015,paszke2017automatic} and the PyTorch-Geometric~\cite{Fey/Lenssen/2019} Python libraries. The neighborhood of each center atom is selected by the nearest-neighbors algorithm implemented in the Scikit-Learn~\cite{scikit-learn} library. The periodic boundary condition and the minimum image convention are enforced in the construction of the neighbor graph and edge values. In feature engineering, we choose $L=40$ and $M=4$ to create radial features for describing the interactions between linked atoms.

Neural networks with more than one hidden layers are more expressive, thus they are more capable of learning the representation of data~\cite{xu2018powerful}. For the GCN0 layer, we use 3 hidden layers, each with 256 hidden units. The number of hidden units are 256 for GCN $i$ $(i = 1, 2, \dots, N-1)$ layers and 1 for GCN $N$ layer. Except for GCN $N$, each hidden layer is followed by a non-linear activation function. It must be emphasized that if too many GCN layers are used, GNN will suffer from over-smoothing~\cite{li2018deeper}, which means the hidden states on each node will be similar because each node in the graph contains the information of all the other nodes due to the aggregation operation of GCN. With residual connections alleviating the problem~\cite{li2019deepgcns}, we can train deeper GCNs. We find the performance of our model is limited if fewer GCNs layers are utilized, since they cannot see “far enough”. In theory an infinitely wide 2 layer network that has a very large cutoff radius should be sufficiently expressive, but in practice it is much more practical, efficient, and effective to train a network with more layers and a smaller cutoff radius. Therefore, in our model, 8 GCN layers with residual connection are used to construct the MDGNN framework.

The loss function of the model is defined as the summation of the mean absolute errors (MAEs) of energy and force,
\begin{equation}
    \begin{aligned}
        \mathcal L & =\gamma\mathcal L_E+(1-\gamma)\mathcal L_F,\\
        \mathcal L_E &  =\frac {1} {\mathscr{N}_\mathrm{atoms}} \sum_{i=1}^{\mathscr{N}_{\mathrm{atoms}}} |E_i-\hat E_{i}|,\\
       \mathcal L_F & = \frac {1} {3\mathscr N_\mathrm{atoms}} \sum_{i=1}^{\mathscr{N}_{\mathrm{atoms}}}\sum_{\alpha=x, y, z} |F_{i\alpha}-\hat F_{i\alpha}| ,
    \end{aligned}
\end{equation}
where $\mathscr N_\mathrm{atoms}$ is the number of the atoms. $E_i$ and $F_i$ are the energies and forces predicted by MDGNN, while $\hat E_i$ and $\hat F_i$ are the true energies and forces. The subscript character $\alpha$ is the Cartesian index and $\gamma$ is the weight coefficient to adjust the relative contributions of the two loss terms during the training process, which is a hyper-parameter that should be set manually. Here we find that with $\gamma=0.2$ and the units of the energies and forces taken as eV and eV/\AA~ respectively, the loss of the MDGNN will converge rapidly.

Because the forces are obtained as derivatives of the predicted potential energy as well as inputs to our loss function, the operations in MDGNN must be doubly-differentiable (having a non-trivial second order derivative) in order to perform back-propagation to train the network. For this reason, we use the double-differentiable function Softplus $\zeta(x)=\log(1+\exp(x))$~\cite{hahnloser2000digital,hahnloser2001permitted,glorot2011deep} as the non-linear activation function. The network is trained with the Adam~\cite{kingma2014adam} optimizer.

To demonstrate the performance of MDGNN on both molecular and crystal structures, we will first compare the performance of SchNet~\cite{schutt2017schnet} and MDGNN with MD17, a public molecular dynamics dataset of eight small molecules. We then compare MDGNN with DeePMD and TeaNet with two crystals, copper and silicon dioxide. Finally, we demonstrate the transferability of MDGNN with aluminum-copper compounds.

\subsection{Molecular and crystal dataset}

The MD17 dataset~\cite{chmiela2017machine} is a molecular dynamics dataset for small molecules. All trajectories are calculated at a temperature of 500 K and a time resolution of 0.5 fs. The potential energy and force labels are computed with PBE+vdW-TS method. The goal of this benchmark is to demonstrate the performance of MDGNN model on molecular systems. Separate MDGNN models are trained for each dataset that is randomly split into a 1000-frame training set, a 1000-frame validation set and the rest frames are the test set. The learning rate is set as $10^{-4}$. 15 \AA~ vacuum layers are inserted in each dimension for all the molecules and the truncated radius is set as 9 \AA~ to construct complete graphs. With $L$ and $M$ set as 40 and 4 respectively, the final results are shown in Table \ref{tab:1}, and 56.25\% of the results obtained by MDGNN outperform that obtained by SchNet~\cite{schutt2017schnet}.


\begin{table}[]
    \caption{\label{tab:1} Comparison between the MAEs of SchNet and MDGNN trained on MD17 datasets using 1000 training samples (energies in meV/atom and forces in meV/\AA). The values in bold represent outperformance on the same task.}
    \centering
    \begin{threeparttable}
    \begin{tabular}{cc|rr} 
        \hline\hline
        ~ & \multicolumn{1}{c|}{~} & \multicolumn{1}{c}{\textbf{SchNet\tnote{$\dagger$}}}& \multicolumn{1}{c}{\textbf{MDGNN}}\\
        \midrule[1pt]
        \multicolumn{1}{c|}{\multirow{2}{*}{Benzene}} & \multicolumn{1}{c|}{energy} & 3.44 & \textbf{1.36}\\
        \cline{2-4}
        \multicolumn{1}{c|}{~} & \multicolumn{1}{c|}{force} & \textbf{13.32} & 14.76\\
        \hline
        \multicolumn{1}{c|}{\multirow{2}{*}{Toluene}} & \multicolumn{1}{c|}{energy} & 5.16 & \textbf{3.92}\\
        \cline{2-4}
        \multicolumn{1}{c|}{~} & \multicolumn{1}{c|}{force} & \textbf{24.51} & 32.89\\ 
        \hline
        \multicolumn{1}{c|}{\multirow{2}{*}{Malonaldehyde}} & \multicolumn{1}{c|}{energy} & 5.59 & \textbf{2.80}\\
        \cline{2-4}
        \multicolumn{1}{c|}{~} & \multicolumn{1}{c|}{force} & \textbf{28.38} & 47.68\\
        \hline
        \multicolumn{1}{c|}{\multirow{2}{*}{Salicylic acid}} & \multicolumn{1}{c|}{energy} & 8.60 & \textbf{7.04}\\
        \cline{2-4}
        \multicolumn{1}{c|}{~} & \multicolumn{1}{c|}{force} & 36.55 & \textbf{36.46}\\
        \hline
        \multicolumn{1}{c|}{\multirow{2}{*}{Aspirin}} & \multicolumn{1}{c|}{energy} & 15.91 & \textbf{6.85}\\
        \cline{2-4}
        \multicolumn{1}{c|}{~} & \multicolumn{1}{c|}{force} & 58.05 & \textbf{57.74}\\
        \hline
        \multicolumn{1}{c|}{\multirow{2}{*}{Ethanol}} & \multicolumn{1}{c|}{energy} & \textbf{3.44} & 5.88\\
        \cline{2-4}
        \multicolumn{1}{c|}{~} & \multicolumn{1}{c|}{force} & \textbf{15.05} & 32.18\\
        \hline
        \multicolumn{1}{c|}{\multirow{2}{*}{Uracil}} & \multicolumn{1}{c|}{energy} & 6.02 & \textbf{4.00}\\
        \cline{2-4}
        \multicolumn{1}{c|}{~} & \multicolumn{1}{c|}{force} & \textbf{24.08} & 37.56\\
        \hline
        \multicolumn{1}{c|}{\multirow{2}{*}{Naphtalene}} & \multicolumn{1}{c|}{energy} & 6.88 & \textbf{4.31}\\
        \cline{2-4}
        \multicolumn{1}{c|}{~} & \multicolumn{1}{c|}{force} & \textbf{24.94} & 37.28\\
        \hline\hline
    \end{tabular}
    \begin{tablenotes}
    \footnotesize
    \item[$\dagger$] The data of SchNet are extracted from~\cite{schutt2017schnet}
    \end{tablenotes}
    \end{threeparttable}
\end{table}

While most proposed machine-learning based methods are solely suitable for small molecules, MDGNN constructs force fields for both molecules and crystals. To demonstrate the performance of MDGNN on crystals, we compare our results of copper (Cu) and silicon dioxide (SiO$_2$) with those from DeePMD and TeaNet. Cu dataset is randomly shuffled and split into training set and test set with the ratio 9:1 while SiO$_2$ dataset is randomly shuffled and split into training set, validation set and test set with the ratio 7:1.5:1.5. The learning rates are set as 1e$^{-4}$ and the cutoff radius of MDGNN is set to 4.0 \AA~for both Cu and SiO$_2$. With $L$ and $M$ set as 40 and 4 respectively, the MAEs of these models are shown in Table \ref{tab:2}. MDGNN outperforms DeepPot-SE (DeePMD) and TeaNet on crystal structures except for the energy MAE of Cu. Although the energy MAE of Cu of MDGNN is larger than that of DeePMD, the error is still far below the chemical accuracy. 

\begin{table}[]
    \caption{\label{tab:2} Comparison about the MAEs of DeepPot-SE (DeePMD)~\cite{zhang2018endtoend} and MDGNN trained on Cu dataset split into training set and test set with the ratio 90\% and 10\%, and the MAEs of TeaNet~\cite{takamoto2019teanet} and MDGNN trained on SiO$_2$ dataset (energies in meV/atom and forces in meV/\AA). The Cu dataset is the public dataset from DeePMD, and the SiO$_2$ dataset is a dataset obtained by first-principles MD simulation, similar to the dataset from TeaNet. For clarity, the better results are indicated in bold.}
    \centering
    \begin{threeparttable}
    \begin{tabular}{ccrrr} 
        \hline\hline
        ~ & \multicolumn{1}{c|}{~} & \multicolumn{1}{c}{\textbf{DeepPot-SE (DeePMD)}} & \multicolumn{1}{c}{\textbf{TeaNet}} & \multicolumn{1}{c}{\textbf{MDGNN}}\\
        \midrule[1pt]
        \multicolumn{1}{c|}{\multirow{2}{*}{Cu}} & \multicolumn{1}{c|}{energy} & \textbf{0.18 (0.25)}\tnote{a} & - & 0.68\\
        \cline{2-5}
        \multicolumn{1}{c|}{~} & \multicolumn{1}{c|}{force} & 90 (90)\tnote{a} & - & \textbf{79.30}\\
        \hline
        \multicolumn{1}{c|}{\multirow{2}{*}{SiO$_2$}} & \multicolumn{1}{c|}{energy} & - & 19.3\tnote{b} & \textbf{0.63}\\
        \cline{2-5}
        \multicolumn{1}{c|}{~} & \multicolumn{1}{c|}{force} & - & 142\tnote{b} & \textbf{69.71}\\ 
        \hline\hline
    \end{tabular}
    \begin{tablenotes}
    \footnotesize
    \item[a] The data of DeepPot-SE and DeePMD are extracted from~\cite{zhang2018endtoend}
    \item[b] The data of TeaNet are extracted from~\cite{takamoto2019teanet}
    \end{tablenotes}
    \end{threeparttable}
\end{table}

\subsection{Aluminum-Copper Compounds}

\begin{figure}
    \centering
    \includegraphics[width=0.48\textwidth]{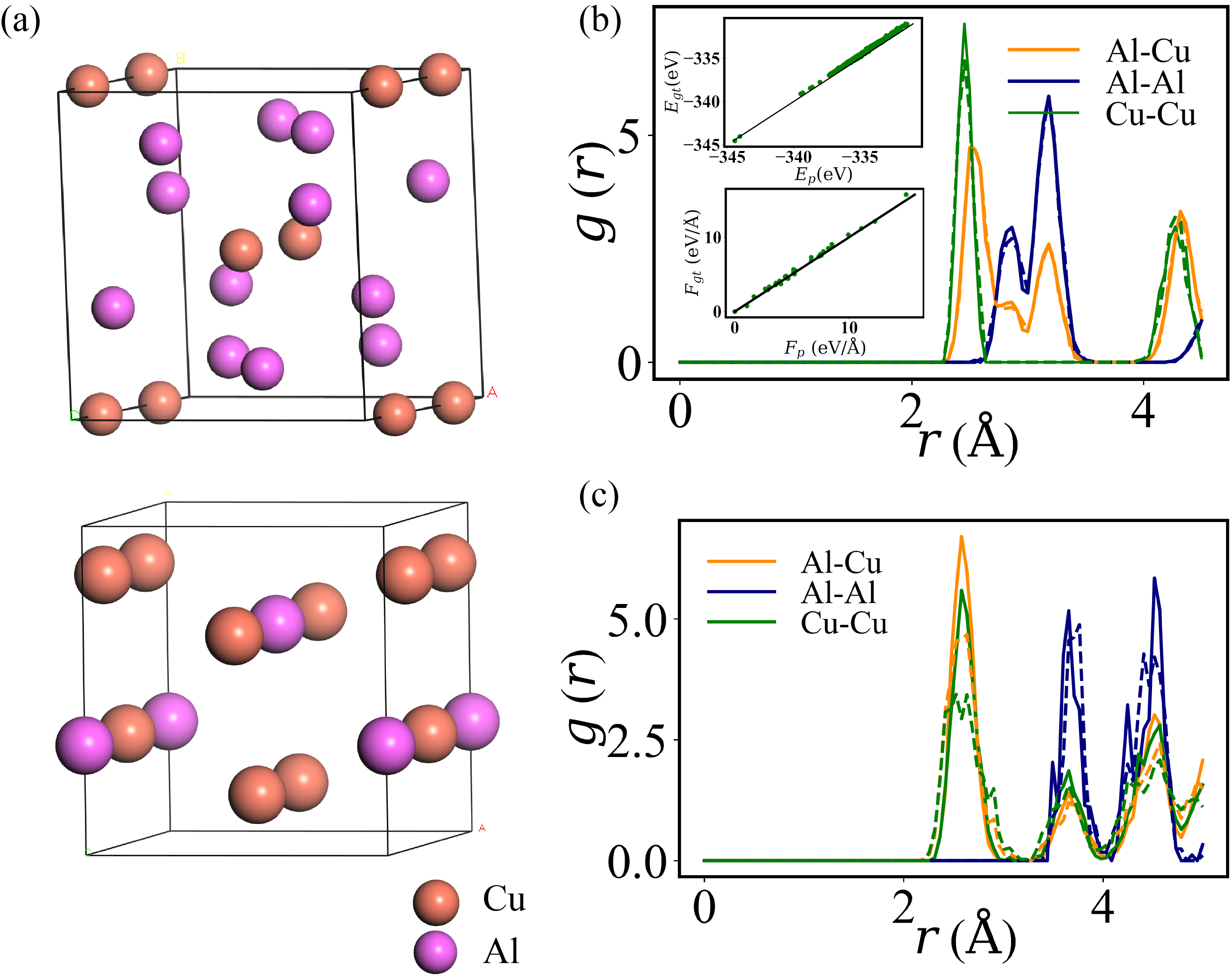}
    \caption{(a) The unit cells of Al$_2$Cu (upper panel) and AlCu$_3$ (lower panel). (b) RDFs of 4 $\times$ 4 $\times$ 4 Al$_2$Cu supercell from MDGNN-based (dotted lines) and ADP force fields (solid lines) under 300 K. The insets are the energy $E$ (top) and forces $F$ (bottom) of Al$_2$Cu test set predicted by the MDGNN {\it v.s.} the ground truth, where the subscripts {\it gt} and {\it p} refer to the ground truth and the prediction, respectively. (c) RDFs of 3 $\times$ 3 $\times$ 3 AlCu$_3$ supercell from MDGNN (dotted lines) trained with Al$_2$Cu dataset and from ADP force fields (solid lines) under 300 K.}
    \label{fig:2}
\end{figure}

Herein we consider two types of Aluminum-Copper compounds: Al$_2$Cu~\cite{bradley1933x} and AlCu$_3$~\cite{tarora1979transformation} [see Fig. \ref{fig:2} (a)]. We will train MDGNN with Al$_2$Cu and test the force fields on AlCu$_3$, for the purpose of demonstrating the transferability. This is a nontrivial task since these two compounds have very different local chemical environments.


Al$_2$Cu has a tetragonal lattice with the I4/mcm space group. The dataset is obtained by classical MD simulations at temperatures from 50 K to 1000 K on a 3 $\times$ 3 $\times$ 2 supercell with angular dependent potential (ADP)~\cite{becker2013considerations,hale2018evaluating} force fields. The cutoff radius of MDGNN is set to 4.5 \AA. The batch size and the learning rate are set as 16 and ${10}^{-4}$, respectively. After 7000 training epochs, the training losses for the energy and force are 1.58 meV/atom and 28.80 meV/\AA~ with corresponding validation losses being 1.98 meV/atom and 28.74 meV/\AA~ respectively. The MAEs of the unseen test set are 1.46 meV/atom and 28.75 meV/\AA. The comparison of the ground truth and the prediction of the potential energies and forces on test dataset is shown in the insets of Fig. \ref{fig:2} (b). The results from MDGNN are in satisfactory agreement with those of the classical MD simulations. We then run MD simulations for a 4 $\times$ 4 $\times$ 4 Al$_2$Cu supercell at a constant temperature of 300 K with MDGNN-based force fields and ADP force fields. The radial distribution functions (RDF) from the MD simulations, presented in Fig. \ref{fig:2} (b), are almost identical for the two force fields. This clearly indicates that MDGNN-based force fields can be well trained on a system of small size and then applied in MD simulation of larger systems of the same material.

To test the transferability of the MDGNN, we perform classical MD simulations for AlCu$_3$ at 300 K on a 3 $\times$ 3 $\times$ 3 supercell using MDGNN trained with the Al$_2$Cu dataset. The RDFs produced by both MDGNN-based and ADP force fields are presented in Fig. \ref{fig:2} (c). It can be observed that MDGNN reproduces all the RDF peak positions, despite some differences in the detailed RDF shapes and values. The transferability of the model is also tested by calculating the formation energy of an Al vacancy in Al$_2$Cu, which is 5.238 eV from ADP force fields and 4.936 eV from MDGNN. The difference is 5.76\%.

Even compared with empirical force fields used in classical MD simulations, MDGNN-based force field is considerably more economical in computational resources. Table \ref{tab:3} presents the computational cost of MD simulations of 10000 time steps with different force fields on a 4 $\times$ 4 $\times$ 4 Al$_2$Cu supercell on the same Intel(R) Xeon(R) CPU E5-2676 v3 CPU. MD with MDGNN-based force field is much faster than embedded atom method (EAM)~\cite{daw1983semiempirical,daw1984embedded,liu1999new} and ADP force fields. We note that MDGNN is trained with ADP force field that contains angle information of different bonds. In contrast, EAM force fields do not take into account such angle information. Another advantage of MDGNN is that contemporary machine learning technology heavily and naturally utilizes GPU, which is significantly faster than CPU for many linear algebra operations. With an Nvidia RTX 2080Ti card, MDGNN runs 27 times faster than ADP force field on CPU, as shown in Table \ref{tab:3}.

\section{Discussion}

\begin{figure*}
    \centering
    \includegraphics[width=1.0\textwidth]{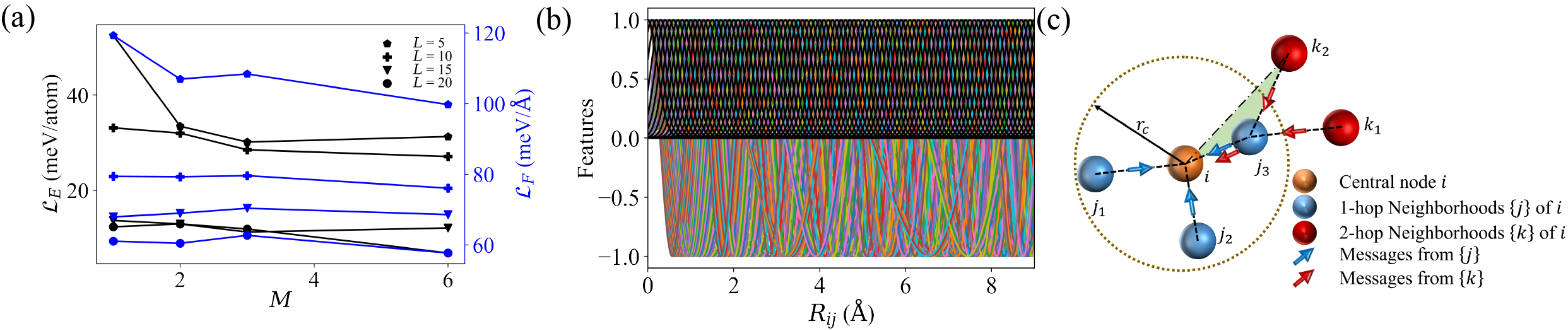}
    \caption{(a) The results of different feature engineerings on aspirin dataset with training size set as 1000. The black lines with different markers represent the energy loss and the blue ones represent forces loss. (b) Every color line represents a corresponding non-linear sine or cosine transformation with some $l\in \{1, 2, \dots, 50\}$ and $m\in \{1, 2, 3\}$. The black lines are the exponential transformations of distances. (c) After a single GCN layer, the messages from 1-hop neighborhoods will be aggregated to the central node, simultaneously these 1-hop neighborhoods will receive the messages of their 1-hop neighborhoods. In the next GCN layer, these 2-hop messages will be sent to the central node and so on. Finally, after $k$ GCN layers, each central node will receive messages from k-hop neighborhoods.}
    \label{fig:3}
\end{figure*}

As explained in Sec. \ref{sec:architecture}, the input features of MDGNN explicitly include high order terms of distances between atoms. In addition, all the input features take values between $0$ and $1$ and are thus automatically normalized. Here we test the performance of MDGNN for different numbers of input features, which is controlled by the choice of $L$ and $M$. As shown in Fig. \ref{fig:3} (a), the choice of $L$ is critical for the accuracy of MDGNN, while the value of $M$ plays a less relevant role. Therefore, we choose $L=40$ and $M=4$ for all calculations in this work.

Compared with the Gaussian-type input features in most of the existing frameworks, our data preprocessing method might be more effective. As shown in Fig. \ref{fig:3} (b), $300$ black lines represent the input features in SchNet~\cite{schutt2017schnet} $\exp\left(-\gamma\parallel{R_{ij}-\mu_k}\parallel^2\right)$, where $\mu_k$ takes value from $0$ \AA~to $30$ \AA~with a $0.1$ \AA~step, and $\gamma = 10$ \AA. While, $300$ color lines represent input features of MDGNN with $L=50$ and $M=3$. Because $\exp(-x^2)$ is a rapidly decreasing function of $x$, most of the input features of SchNet are zero for any specific $R_{ij}$. In contrast, most of the sine- and cosine-type input features are finite for any $R_{ij}$. We suspect our input features provide more information for the neural network to construct the force field.

Force fields not only depend on distances between atoms, but also depend on angular information such as bond angles. Indeed, except for Cu, we have been training MDGNN with either classical force fields with angle information or first principles MD data. The excellent accuracy indicates MDGNN is able to capture the angle information in the data, despite the fact that the only input features are distances between atoms. Due to the sophisticated multilayer nonlinear structure, deep neural networks are nontransparent and their predictions are not traceable by humans. We partially interpret the working mechanism of MDGNN in Fig. \ref{fig:3} (c). Although each node aggregates the embeddings of only its neighbors in a single GCN layer, the next GCN layer will extract the information concerning the second-order neighbors from its first-order neighbors. Similarly, with $k$ GCN layers, each node in the graph will aggregate $k$-hop information. For three nodes linked with each other in a graph, forming a triangle, MDGNN may learn the angular information implicitly from the lengths of the three sides of the triangle. 

\begin{table}[b]
\caption{\label{tab:3}
Comparison of computational cost of 10000 MD time steps with different force fields on a 4 $\times$ 4 $\times$ 4 Al$_2$Cu supercell. Note that the MD simulations with angular dependent potential (ADP) and embedded atom method (EAM) force fields could be run only on CPU in atomic simulation environment of python. The results from empirical force fields and MDGNN (CPU) are obtained on the same CPU. The result of MDGNN (GPU) is obtained with the model allocated on a GPU.}
\begin{ruledtabular}
\begin{tabular}{ccccc}
 Calculator & EAM & ADP & MDGNN & MDGNN (GPU)  \\
\hline
Time (hour) & 5.10 & 11.35 & 3.24 & 0.61  \\
\end{tabular}
\end{ruledtabular}
\end{table}

There are also several other GNN-based architectures that could be used in molecular dynamics. For example, visual interaction networks~\cite{watters2017visual}, a model for the dynamics of the physical systems, learns the interactions from raw frames of motion. If the atoms are regarded as hard spheres, this network might be useful in learning the interactions between atoms. However, considering the down-sampling of the convolutional neural networks, the visual interaction networks could not be directly applied to the MD simulations. Alternatively, the data in molecular dynamics simulations could also be viewed as point clouds, which could be processed by PointNet~\cite{qi2017pointnet} and PointNet++~\cite{qi2017pointnet++} with some modifications to preserve all the symmetries. To sum up, these proposed architectures should be capable of processing dataset of MD simulations if certain symmetry-adapted tricks are introduced.

We introduce a symmetry-adapted graph neural network architecture to construct molecular dynamics force fields for both molecules and crystals. Previously proposed non-GNN-based models are mostly based on traditional fully connected neural networks, for which preprocessing of input data is required to ensure the symmetries such as the translation, rotation and permutation invariance. On the other hand, most of the proposed GNN-based models are solely applied to molecular structures. MDGNN introduced here is based on graph convolutional networks, which can naturally handle graph-structured data. This framework transforms both molecular and crystal structures to crystal graphs with periodic boundary condition and minimum image convention. Force fields constructed by MDGNN can accurately describe systems which are larger than the training system. For materials with the same type of atoms as the training material, MDGNN can capture the most prominent features in the radial distribution function. Although the input features of MDGNN do not explicitly contain the bond angle information of atoms, the results demonstrate that this information is captured and utilized in the MD simulations with MDGNN-based force fields. MDGNN has significant potential to become a tool for constructing accurate force fields and potential energy surfaces with low computational cost and good transferability for different systems.
\section{Methods}

The classical MD simulation in the canonical NVT ensemble is performed with the python library atomic simulation environment~\cite{ase-paper,ISI:000175131400009}. The Langevin dynamics of aluminum-copper compound is run with the angular dependent potential~\cite{apostol2011interatomic} force field. The initial momenta is set to a Maxwell-Boltzmann distribution with the temperature set to 50 K. The desired temperature, time step and friction term are set to 1000 K, 1 fs and 0.2 atomic unit, respectively. We obtained 7000 frames of the canonical ensemble simulations as dataset. Constant temperature simulations at 300 K for a larger Al$_2$Cu supercell and AlCu$_3$ are also performed to compare with MDGNN results.

The dataset of SiO$_2$ is generated with first-principles MD simulation which is performed with Vienna \emph{ab initio} Simulation Package~\cite{kresse1993ab,kresse1994ab,kresse1996efficiency,kresse1996efficient} using the projector-augmented wave~\cite{blochl1994projector,kresse1999ultrasoft} pseudopotentials. Generalized gradient approximation is included in the Perdew-Berke-Ernzerhof exchange-correlation potential~\cite{perdew1996generalized,perdew78ernzerhof}. The cutoff of plane waves is 400 eV and the Monkhorst-Packk-point meshes of 3 $\times$ 3 $\times$ 3 are adopted for MD simulations. The simulation is carried out under the NVT ensemble with the temperature increasing from 300 K to 1000 K. With the time step of 1 fs, 7000 frames are obtained from the first-principles MD simulation as the dataset.

\section{Data Availability}
The raw data of MD17 is available at \url{http://quantum-machine.org/datasets/}. The raw data of Cu is available at \url{http://www.deepmd.org/database/deeppot-se-data/}.

\section{Code Availability}
The implementations of MDGNN described in the paper will be available after the manuscript is accepted for publication.


\bibliographystyle{naturemag}
\bibliography{ref}

\begin{acknowledgements}
This work was supported by the Basic Science Center Project of NSFC (Grant No. 51788104), the Ministry of Science and Technology of China (Grant Nos. 2016YFA0301001 and 2017YFB0701502), and the Beijing Advanced Innovation Center for Materials Genome Engineering.

\end{acknowledgements}

\section{AUTHOR CONTRIBUTIONS}
Z.W., C.W. and W.D. conceived the idea, designed the research. Z.W. implemented the codes and S.Z. further optimized these codes. Z.W. and S.D. performed the molecular dynamics simulations. W.D., Y.X. and B.G. supervised the work. All authors discussed the results and were involved in the writing of the manuscript.

\section{COMPETING INTERESTS}
The authors declare no competing interests.

\end{document}